\documentclass[prl,twocolumn,superscriptaddress,showpacs,amsmath,amssymb,letterpaper,floatfix]{revtex4}

\usepackage{graphicx}
\usepackage{dcolumn}
\usepackage{bm}
\usepackage{color}

\graphicspath{{./}}

\def\DE {\Delta E}
\def\Mbc {M_{\rm bc}}

\begin{document}

\preprint{\vbox{ \hbox{   }
                 \hbox{KEK Preprint 2007-}
                 \hbox{Belle Preprint 2007-}
}}

\title{
 Search for $B^0 \to J/\psi \phi $
  Decays }

\affiliation{Budker Institute of Nuclear Physics, Novosibirsk}
\affiliation{University of Cincinnati, Cincinnati, Ohio 45221}
\affiliation{Justus-Liebig-Universit\"at Gie\ss{}en, Gie\ss{}en}
\affiliation{The Graduate University for Advanced Studies, Hayama}
\affiliation{Hanyang University, Seoul}
\affiliation{University of Hawaii, Honolulu, Hawaii 96822}
\affiliation{High Energy Accelerator Research Organization (KEK), Tsukuba}
\affiliation{Hiroshima Institute of Technology, Hiroshima}
\affiliation{Institute of High Energy Physics, Chinese Academy of Sciences, Beijing}
\affiliation{Institute of High Energy Physics, Vienna}
\affiliation{Institute of High Energy Physics, Protvino}
\affiliation{Institute for Theoretical and Experimental Physics, Moscow}
\affiliation{J. Stefan Institute, Ljubljana}
\affiliation{Kanagawa University, Yokohama}
\affiliation{Korea University, Seoul}
\affiliation{Kyungpook National University, Taegu}
\affiliation{\'Ecole Polytechnique F\'ed\'erale de Lausanne (EPFL), Lausanne}
\affiliation{Faculty of Mathematics and Physics, University of Ljubljana, Ljubljana}
\affiliation{University of Maribor, Maribor}
\affiliation{University of Melbourne, School of Physics, Victoria 3010}
\affiliation{Nagoya University, Nagoya}
\affiliation{Nara Women's University, Nara}
\affiliation{National Central University, Chung-li}
\affiliation{National United University, Miao Li}
\affiliation{Department of Physics, National Taiwan University, Taipei}
\affiliation{H. Niewodniczanski Institute of Nuclear Physics, Krakow}
\affiliation{Nippon Dental University, Niigata}
\affiliation{Niigata University, Niigata}
\affiliation{University of Nova Gorica, Nova Gorica}
\affiliation{Osaka City University, Osaka}
\affiliation{Osaka University, Osaka}
\affiliation{Panjab University, Chandigarh}
\affiliation{University of Science and Technology of China, Hefei}
\affiliation{Seoul National University, Seoul}
\affiliation{Sungkyunkwan University, Suwon}
\affiliation{University of Sydney, Sydney, New South Wales}
\affiliation{Tata Institute of Fundamental Research, Mumbai}
\affiliation{Toho University, Funabashi}
\affiliation{Tohoku Gakuin University, Tagajo}
\affiliation{Tohoku University, Sendai}
\affiliation{Department of Physics, University of Tokyo, Tokyo}
\affiliation{Tokyo Institute of Technology, Tokyo}
\affiliation{Tokyo Metropolitan University, Tokyo}
\affiliation{Tokyo University of Agriculture and Technology, Tokyo}
\affiliation{Virginia Polytechnic Institute and State University, Blacksburg, Virginia 24061}
\affiliation{Yonsei University, Seoul}
  \author{Y.~Liu}\affiliation{The Graduate University for Advanced Studies, Hayama} 
  \author{K.~Trabelsi}\affiliation{High Energy Accelerator Research Organization (KEK), Tsukuba} 
   \author{I.~Adachi}\affiliation{High Energy Accelerator Research Organization (KEK), Tsukuba} 
   \author{H.~Aihara}\affiliation{Department of Physics, University of Tokyo, Tokyo} 
 \author{K.~Arinstein}\affiliation{Budker Institute of Nuclear Physics, Novosibirsk} 
   \author{V.~Aulchenko}\affiliation{Budker Institute of Nuclear Physics, Novosibirsk} 
   \author{T.~Aushev}\affiliation{\'Ecole Polytechnique F\'ed\'erale de Lausanne (EPFL), Lausanne}\affiliation{Institute for Theoretical and Experimental Physics, Moscow} 
   \author{T.~Aziz}\affiliation{Tata Institute of Fundamental Research, Mumbai} 
   \author{A.~M.~Bakich}\affiliation{University of Sydney, Sydney, New South Wales} 
   \author{A.~Bay}\affiliation{\'Ecole Polytechnique F\'ed\'erale de Lausanne (EPFL), Lausanne} 
% ...
\collaboration{The Belle Collaboration}


\begin{abstract}
We report
a search for the decay $B^0 \to J/\psi \phi$,
using a sample of 657
$\times\ 10^6$ $B\bar{B}$ pairs collected with the Belle detector at
the $\Upsilon(4S)$ resonance. No statistically significant signal is
found and an upper limit for the
branching fraction is determined to be
$\mathcal{B}(B^0 \to J/\psi \phi) < 9.4 \times\ 10^{-7}$ at
90\% confidence level.
\end{abstract}

 \maketitle

\begin{figure}[hbtp]
\includegraphics[width=0.4\textwidth]{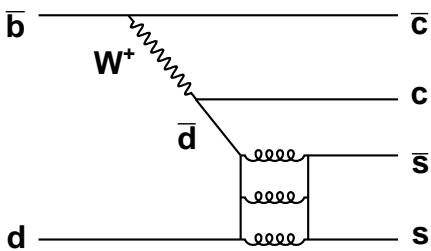}
\caption{ Quark-level diagram for $B^0 \to J/\psi \phi$ decay }
\label{diagram}
\end{figure}

Studies of exclusive $B$ meson decays to charmonium play an
important role in exploring $CP$ violation \cite{charmonium_cpv} and
establishing the Kobayashi-Maskawa anzatz \cite{KM} for $CP$ violation
in the Standard Model.
Such studies have also resulted
in observations of new resonant states that include a $(c\bar c)$
pair \cite{new_state, new_state2, new_state3}.
The decay $B^0 \to J/\psi \phi$
is expected to proceed mainly
via a Cabibbo-suppressed and
color-suppressed transition ($b \to c\bar cd$)
with rescattering, as shown in Fig.~\ref{diagram}.
In $B$ decays, effects presumably due to rescattering have been
seen in various decay processes.
For example, the large branching fractions observed for
$B^0 \to D_s^{-} K^+$ \cite{dsk} and $B^- \to \chi_{c0} K^-$ \cite{chic0k}
decays can be attributed to rescattering processes \cite{block,colangelo}.
An isospin analysis on $B \to DK^{(*)}$ decays indicates significant
final state rescattering effects \cite{dk}.
Final state rescattering may play an important role
in understanding patterns of $CP$ asymmetries in $B$ decays to two
charmless pseudoscalars \cite{b2pp}.
Studies of $B$ decays such as $B^0 \to J/\psi \phi $, which would
proceed mainly via rescattering,
provide useful information for understanding
rescattering mechanisms.
Previously, the BaBar collaboration reported
a search for this decay mode
and set an upper limit for the branching fraction
${\cal B} <  9.2 \times\ 10^{-6}$ at the 90\% confidence level
based on 56 $\times\ 10^6$ $B\bar B$ pairs \cite{jpsiphi}.

In this paper, we report the results of a search for the decay mode
$B^0 \to J/\psi \phi $
using the Belle detector \cite{belle} at the KEKB energy-asymmetric
$e^+e^-$ collider \cite{kekb}
based on a 605 $\rm fb^{-1}$ data sample containing 657 $\times\ 10^6$
$B\bar{B}$ pairs.
This sample is more
than an order of magnitude larger than that used previously.

The Belle detector is a large-solid-angle magnetic spectrometer that
consists of a silicon vertex detector (SVD), a 50-layer central
drift chamber (CDC), an array of aerogel threshold Cherenkov
counters (ACC), a barrel-like arrangement of time-of-flight
scintillation counters (TOF), and an electromagnetic calorimeter
comprised of CsI(Tl) crystals (ECL). These detectors are located
inside a superconducting solenoid coil that provides a 1.5 T
magnetic field. An iron flux-return located outside of the coil is
instrumented to detect $K_L^0$ mesons and to identify muons (KLM).

Events are required to pass a basic hadronic event selection
\cite{PRD03}. To suppress continuum background ($e^+e^- \to q \bar q$,
where $q = u, d, s, c$), we require the ratio of the second to
zeroth Fox-Wolfram moments \cite{FW} to be less than 0.5.

Candidates for $B^0\to\ J/\psi \phi$ decays
are reconstructed from the
decays $J/\psi \to \ell^+ \ell^-$ ($\ell = e, \mu$) and
$\phi \to K^+K^-$.
The selection criteria for the $J/\psi$ decaying to $\ell^+\ell^-$ are
identical to those used in our previous papers \cite{PRD03}.
 $J/\psi$ candidates are pairs of oppositely
charged tracks that originate from a region within 5~cm of the
nominal interaction point (IP) along the beam direction and are
positively identified as leptons. In order to reduce the effect of
bremsstrahlung or final-state radiation, photons detected in the ECL
within $0.05$ radians of the original $e^-$ or $e^+$ direction are
included in the calculation of the $e^+e^-(\gamma)$ invariant mass.
Because of the radiative low-mass tail, the $J/\psi$ candidates are
required to be within an asymmetric invariant mass window:
$-150~(-60)$ $\rm{MeV}$/$c^2$
$<M_{e^+e^-(\gamma)}(M_{\mu^+\mu^-})-m_{J/\psi} <$+36~(+36)
$\rm{MeV}$/$c^2$, where $m_{J/\psi}$ is the nominal $J/\psi$ mass
\cite{pdg}. In order to improve the $J/\psi$ momentum
resolution, a vertex and mass constrained fit to the
reconstructed $J/\psi$ candidates is then performed and a loose cut
on the vertex fit quality is applied.

In order to identify hadrons, a likelihood $L_i$ for each  hadron
type $i$ ($i = \pi, K$ and $p$) is formed using information from the
ACC, the TOF, and $dE/dx$ measurements from the CDC.
Kaons from the $\phi$ meson are selected with the
requirement $L_{K}/(L_{K}+L_{\pi})> 0.7$,
which has an efficiency of 90.0\% and a 5.9\% probability to
misidentify a pion as kaon.
This requirement is chosen to minimize the upper limit expected
in the absence of a real signal, based on
studies of signal and background Monte Carlo (MC) events.
We reconstruct $\phi$ candidates
from pairs of $K^+K^-$ candidates, where we
require the invariant mass to be
within $\pm 10$ MeV/$c^2$ of the nominal $\phi$ mass~\cite{pdg}.

$B^0$ mesons are reconstructed by combining a $J/\psi$ with a $\phi$
candidate. We identify $B^0$ candidates using two
kinematic variables calculated in the center-of-mass system: the
beam-energy constrained mass ($\Mbc \equiv \sqrt{E_{\rm
beam}^{2}-P_{B}^{2}}$) and the energy difference ($\DE \equiv E_{B}-
E_{\rm beam}$), where $E_{\rm{beam}}$ is the beam energy, and $P_{B}$
and $E_{B}$ are the reconstructed momentum and energy of the $B^0$
candidate. We select $B$ candidates within the range
$-0.2\ {\rm GeV} < \DE < 0.3$ $\rm{GeV}$ and
$5.27$ GeV/$c^2$ $< \Mbc < 5.29$ GeV/$c^2$ for the final analysis.
After all selection requirements, about 4.9\%
of the events contain more than one $B^0$ candidate.
For these events, we choose the
$B$ candidate whose daughter particle $\phi$
mass is closest to the nominal value.
Finally, a total of 85 candidates are selected.

The dominant background comes from
$B\bar{B}$ events with $B$ decays to $J/\psi$.
We use a MC sample corresponding to 3.86$\times\ 10^{10}$
generic $B\bar{B}$ decays that includes all known $B \to J/\psi X$ processes to
investigate these backgrounds. We find that the dominant backgrounds come from
$B^0 \to J/\psi K^{*0}(892)[\to K^-\pi^+]$ and
$B^{0/-} \to J/\psi K_1(1270)[\to K^-\pi^+\pi^{0/-}]$~\cite{ksk1}.
In both cases, a pion is misidentified as a kaon, and
in the latter case, the other pion is
missed.  The former has a peak at $\DE \sim 0.1$ GeV,
while the latter has a broad peak in the
negative $\DE$ region.
The remaining background is due to random combinations of $J/\psi$ and $\phi$
candidates and does not peak in the $\DE$ distribution
(referred to as combinatorial background).

The signal yield is extracted by performing an
unbinned extended maximum-likelihood fit to the $\DE$ distribution of
candidate events.
The likelihood function is given as
\begin{equation}
{\cal L} = \frac{e^{-\sum
\limits_{k}N_{k}}}{N!}\prod^N_{i=1}\left[\sum_{k}N_{ k}\times
P_k(\Delta E^{i})\right],
\end{equation}
where $N$ is the total number of candidate events, $i$ is the
identifier of the $i$-th event, $N_{k}$ and $P_{k}$ are the yield
and probability density function (PDF) of the component $k$, which
corresponds to the signal, $J/\psi K_1$, $J/\psi K^{*0}$, and
combinatorial backgrounds.

The signal PDF is modeled by a sum of two Gaussians.
The background PDFs are two Gaussians for the $J/\psi K_1$ component,
a bifurcated Gaussian for the $J/\psi K^{*0}$ component and
a second-order polynomial
for combinatorial background, respectively.
The parameters of these PDFs are determined from MC simulations.
We use $B^0 \to J/\psi K^{*0}$ decay with $K^{*0} \to K^-\pi^+$ as a
control data sample to
correct for small differences between data and MC in the mean and width
of the signal PDF.
The $J/\psi K_1$ component shape is verified by comparing
data and MC events in the  $K^+K^-$ mass sideband region
(1.04 -- 1.10 GeV/$c^2$), while events in the
5.22 GeV/$c^2 < M_{\rm bc} < $ 5.26 GeV/$c^2$ and $K^+K^-$ mass sideband
region are used to check the combinatorial background shape.
Possible differences between data and MC are included in the
systematic errors.

\begin{figure*}[hbtp]
\includegraphics[width=0.6\textwidth]{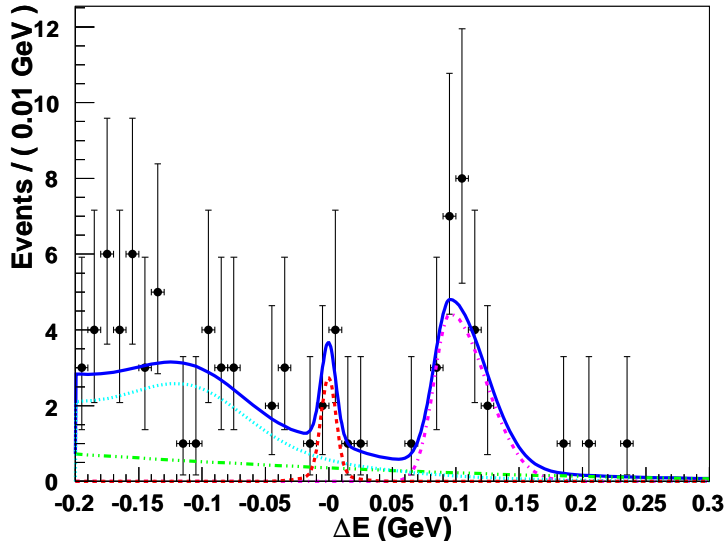}
\caption{ $\DE$ distribution for
$B^0 \to J/\psi \phi$ candidates.
The
curves show the signal
 (red dashed) and the background components
 (cyan dashed for $J/\psi K_1$, magenta dot-dashed for
 $J/\psi K^{*0}$ and green triple-dot dashed for combinatorial)
 as well as the overall fit (blue solid curve).
} \label{sbox}
\end{figure*}

\begin{table}[!hbtp]
\begin{center}
\caption{\label{result}
 Summary of the results:
 upper limits are at the 90\% confidence level.
}
\vspace{2 mm}
\begin{tabular}{cc}
\hline \hline
 Signal yield    & $4.6\ ^{+3.1}_{-2.5}$ \\
 Significance    & $2.3\sigma$    \\
 Upper limit of signal yield ($Y_{90}$) & 9.5  \\
 Detection efficiency ($\epsilon$)  & 26.2\%  \\
 Upper limit of branching fraction  & ~~$< 9.4 \times\ 10^{-7}$ \\
\hline\hline
\end{tabular}
\end{center}
\end{table}

In the fit, all $N_k$ values are free parameters. Figure~\ref{sbox}
shows the $\DE$ distribution of the $B^0 \to J/\psi \phi$ candidates
together with the fit result. We obtain a signal yield of $4.6\
^{+3.1}_{-2.5}$ events with a statistical significance of
$2.3\sigma$. This statistical significance is defined as
$\sqrt{-2\ln({\cal L}_{0}/{\cal L}_{\rm max})}$, where ${\cal
L}_{\rm max}$ and ${\cal L}_{0}$ denote the maximum likelihood with
the fitted signal yield and with the yield fixed to zero,
respectively. The number of misidentified $B^0 \to J/\psi K^{*0}$
events obtained from the fit is $22.5\ ^{+5.4}_{-4.8}$ and is
consistent with the expectation obtained from MC simulation
incorporating the misidentification probability and the world
average branching fraction~\cite{pdg}.

As no significant signal is found for the
$B^0 \to J/\psi \phi$ decay mode,
we obtain an upper limit on the yield at the
90\% confidence level ($Y_{90}$)
by a frequentist method using ensembles of pseudo-experiments.
For a given signal yield, 10000 sets of signal and background events
are generated
according to the PDFs, and fits are performed.  The confidence level
is obtained as the fraction of samples that give a fit yield larger
than that of data (4.6).
We account for systematic error by smearing the fit yield by the
total systematic error described below.
We scan signal yields and obtain $Y_{90} = 9.5$.

The corresponding branching fraction upper limit is determined with
\begin{equation}
{\cal B} ~<~ \frac{Y_{90}}{ \epsilon \times
N_{B\bar{B}} \times {\cal B}(J/\psi \to \ell^+\ell^-) \times {\cal
B}(\phi \to K^+ K^-) }
\label{bf}
\end{equation}
Here
$N_{B\bar{B}}$ is the number of $B\bar{B}$ pairs, and we
use the world averages \cite{pdg} for the branching fractions of
${\cal B}(J/\psi \to \ell^+\ell^-)$ and ${\cal B}(\phi \to K^+ K^-)$.
The efficiency ($\epsilon = 26.2\%$)
is determined from a signal
MC sample with the same selection as used for the data,
where a correction for muon identification efficiency 
due to differences between data and MC is included (described below).
The fractions of
neutral and charged $B$ mesons produced in $\Upsilon(4S)$ decays are
assumed to be equal.
These results are summarized in Table~\ref{result} and
an upper limit at the 90\% confidence level is obtained
\begin{equation}
\mathcal{B}(B^0 \to J/\psi \phi) < 9.4 \times\ 10^{-7}.
\end{equation}

\begin {table}[htp]
\begin {center}
\caption {Summary of systematic uncertainties (\%)
other than signal yield extraction (denominator in Eq. 2).  }
\vspace{2 mm}
\begin {tabular}{cc}
\hline \hline
Uncertainty Source &
Uncertainty (\%) \\
\hline
Tracking efficiency & $4.2$ \\
Lepton ID efficiency & $4.2$   \\
Kaon ID efficiency & $2.2$   \\
Polarization  & 2.6  \\
$J/\psi$ branching fractions & $1.0$\\
$\phi$ branching fraction   & $1.2$\\
Number of $B\bar B$   & $1.4$\\
\hline
Total & $7.2$\\
\hline
\hline
\end {tabular}
\label{sys-err}
\end {center}
\end {table}

\begin {table}[htp]
\begin {center}
\caption {Summary of systematic uncertainties on signal yield
 ($\rm{\Delta} \it n $) by source.
}
\vspace{2 mm}
\begin {tabular}{ccc}
\hline \hline
Uncertainty Source & $(+\sigma)\rm{\Delta} \it n $ & $(-\sigma)\rm{\Delta} \it n$ \\
\hline
$K_1(1270)$ & $1.0$ & $1.2$   \\
$K^{*0}$ & $< 0.1$ & $< 0.1$  \\
Combinatorial background & 0.1  & 0.2 \\
Signal  & $< 0.1$ & $0.1$\\
\hline
Total & $1.0$& $1.2$\\
\hline \hline
\end {tabular}
\label{sys-err2}
\end {center}
\end {table}

The sources and sizes of systematic uncertainties are summarized in
Tables~\ref{sys-err} and \ref{sys-err2}. The dominant sources of
systematic error in the reconstruction efficiency are tracking
efficiency and particle identification.  Uncertainties in the
tracking efficiency are estimated by linearly summing the
momentum-dependent single track systematic errors ($\sim 1\%$ per
track). We use control samples of $J/\psi \to \ell^+\ell^-$ and
$e^+e^- \to e^+e^- \ell^+\ell^-$ events to estimate lepton
identification efficiency corrections and uncertainties. For the
$J/\psi \to \mu^+\mu^-$ mode, we find the efficiency for a muon
track in the data to be $(4.3 \pm 3.1)\%$ lower than that of MC
simulation. We correct the efficiency for this difference and assign
a 3.1\% uncertainty per muon track. For the $J/\psi \to e^+e^-$
mode, the difference between efficiencies in the data and in the MC
simulation is small, and we assign a 2.7\% uncertainty per electron
track based on their difference and errors. We assign an uncertainty
of 1.2\% per kaon track, which is obtained using kinematically
identified kaons in a $D^{*+} \to D^0\pi^+ [D^0 \to K^-\pi^+]$
sample. Because of the small energy release in $\phi \to K^+K^-$
decay, the selection efficiency of $B^0 \to J/\psi \phi$ decays
depends only weakly on the final state polarization. We use an
average of the efficiencies for fully longitudinally and
transversely polarized cases and assign the difference as a
systematic error ($\pm 2.6\%$ including MC statistical error).
The systematic errors
due to signal and background shapes are evaluated by varying each of
the PDF parameters
by its uncertainty.
We find that the $J/\psi K_1$ component uncertainty is dominant
and that the total systematic error on the signal yield is
$+21.7\%/-26.1\%$ (Table~\ref{sys-err2}).
Adding all sources in quadrature
and conservatively taking the larger of the asymmetric errors,
the total systematic error is estimated to be 27\%.
As a cross check of the MC efficiency and analysis procedure, we
apply the same analysis procedure to the $B^0 \to J/\psi K^{*0}$
control sample and obtain ${\cal B} = (1.24 \pm 0.01) \times
10^{-3}$ (the error is statistical only). This is consistent with
the world average~\cite{pdg} within its uncertainty and the
estimated systematic error of the efficiency mentioned above.

In summary, we have searched for $B^0 \to J/\psi \phi$ decays.
No statistically significant signal
is found and an upper limit for this decay
is determined to be $\mathcal{B}(B^0 \to
J/\psi \phi) < 9.4 \times\ 10^{-7}$ at the 90\% confidence level.
This result improves upon the previous result~\cite{jpsiphi} by
 about a factor of 10 and imposes a more stringent constraint on rescattering effects in
$B^0 \to J/\psi \phi$ decays.

\begin{acknowledgments}
We thank the KEKB group for their excellent operation of the
accelerator, the KEK cryogenics group for their efficient
operation of the solenoid, and the KEK computer group and
the National Institute of Informatics for valuable computing
and SINET3 network support. We acknowledge support from
the Ministry of Education, Culture, Sports, Science, and
Technology of Japan and the Japan Society for the Promotion
of Science; the Australian Research Council and the
Australian Department of Education, Science and Training;
the National Natural Science Foundation of China under
contract No.~10575109 and 10775142; the Department of
Science and Technology of India;
the BK21 program of the Ministry of Education of Korea,
the CHEP SRC program and Basic Research program
(grant No.~R01-2005-000-10089-0) of the Korea Science and
Engineering Foundation, and the Pure Basic Research Group
program of the Korea Research Foundation;
the Polish State Committee for Scientific Research;
the Ministry of Education and Science of the Russian
Federation and the Russian Federal Agency for Atomic Energy;
the Slovenian Research Agency;  the Swiss
National Science Foundation; the National Science Council
and the Ministry of Education of Taiwan; and the U.S.\
Department of Energy.
\end{acknowledgments}


\begin{thebibliography}{999}
\bibitem{ksk1}
Hereafter, we denote $K^{*0}(892)$ and $K_1(1270)$ as simply $K^{*0}$
and $K_1$, respectively.
Also, inclusion of the charge-conjugate states is implied throughout
this paper.
\end{thebibliography}
\end{document}